\documentclass[twocolumn,showpacs,pra,floatfix]{revtex4}
\usepackage{amsmath}
\usepackage{amssymb}
\usepackage{epsfig}
\begin{document}
\title{Quantum Chernoff bound as a measure of nonclassicality 
for one-mode Gaussian states}
\author{M\u ad\u alina Boca}
\author{Iulia Ghiu}
\author{Paulina Marian}  
\author{ Tudor A. Marian\footnote{email: tudor.marian@g.unibuc.ro}}
\affiliation{ Centre for Advanced  Quantum Physics,
University of Bucharest, P.O.Box MG-11,
R-077125 Bucharest-M\u{a}gurele, Romania}
\date{\today}

\begin{abstract}
We evaluate a Gaussian distance-type degree of nonclassicality 
for a single-mode Gaussian state of the quantum radiation field 
by use of the recently discovered quantum Chernoff bound. 
The general properties of the quantum Chernoff overlap and 
its relation to the Uhlmann fidelity are interestingly illustrated 
by our approach. 
\end{abstract}
\maketitle
Nonclassicality of a multimode state of the quantum radiation field 
is usually identified by inspection of its diagonal $P$ representation. 
States possessing a well-behaved $P$ representation are termed 
classical \cite{TG}. On the contrary, a negative or a highly 
singular $P$ representation (i.e., more singular than  Dirac's 
$\delta$) characterizes a nonclassical state. 
A quantitative measure of nonclassicality was first proposed 
by Hillery \cite{Hi} as a properly defined  distance  between 
the given nonclassical state and the convex set of all classical 
states. However, the trace metric employed in Refs.\cite{Hi} 
turned out to be difficult to deal with analytically. Therefore,
the original definition of a {\em nonclassical distance} 
was subsequently modified twofold: first, by restricting the set
of all classical states to a tractable subset identified 
by a classicality criterion, and second, by using more convenient 
distances between Gaussian states. Specifically, we mention 
the Hilbert-Schmidt \cite{HS} and the Bures metric \cite{PTH1,PTH2}, 
as well as the relative-entropy measure in Refs.\cite{PTH1,PTH2}. 
Note also the recent work on defining a distance-type polarization 
degree in Refs.\cite{Soto}. As shown in Refs.\cite{PTH1,PTH2}, 
any good distance-type measure of nonclassicality has to satisfy 
several requirements, which proved to be of both principled 
and practical importance. The Bures metric \cite{Bu} singled out 
among the proposed distance measures by its conspicuous
distinguishability features \cite{Fuchs} and its complete 
agreement with the Lee's nonclassical depth \cite{CTL}.

The present work parallels previous papers \cite{PTH1,PTH2} 
in studying nonclassicality of one-mode Gaussian states 
by use of distance-type measures with remarkable distinguishability 
virtues. We here propose a nonclassicality measure built with 
the quantum Chernoff bound \cite{ch1,ch2}, whose main properties 
we briefly recall in what follows. In the symmetric quantum 
hypothesis, let $\rho$ and $\sigma$ be two equiprobable states 
of a quantum system. We denote by $P_{min}^{(n)}(\rho, \sigma)$ 
the minimal error probability of discriminating these states 
in $n$ independent tests on identical copies of the system, 
all of them prepared in the same state, which is either $\rho$ or $\sigma$. 
The optimal asymptotic testing $(n\rightarrow\infty)$ leads 
to an upper bound called the quantum Chernoff bound,
\begin{equation}
\xi_{QCB}(\rho,\sigma):=-\lim_{n\rightarrow\infty}\left\{\frac
{1}{n}\ln \left [P_{min}^{(n)}(\rho, \sigma)\right ]\right \}.
\end{equation}
Quite recently, it has been proven an intrinsic formula for this bound
\cite{ch1,ch2}: 
\begin{equation}
\xi_{QCB}(\rho,\sigma)=-\ln\left[\min_{0\leq s \leq 1}
{\rm Tr}(\rho^s \sigma^{1-s})\right]. \label{QCB}
\end{equation}
The question of finding the above minimum is the quantum  counterpart of a classical Bayesian 
probability  problem formulated and solved long ago 
by Herman Chernoff \cite{Cher}. The quantities
\begin{equation}
Q_s(\rho,\sigma):={\rm Tr}(\rho^s \sigma^{1-s}) \label{Renyi}
\end{equation}
are the quantum analogues of the classical R\'enyi overlaps 
discussed in Ref.\cite{Fuchs} as being distinguishability 
measures in their own right. According to Eq.\ (\ref{QCB}),
their minimum determines the quantum Chernoff bound:
\begin{equation}
Q(\rho,\sigma):=\min_{0\leq s \leq 1}Q_s(\rho,\sigma)
=\exp{\left[-\xi_{QCB}(\rho,\sigma)\right].} \label{Q}
\end{equation}
In what follows, the non-negative function $Q(\rho, \sigma )$ 
defined in Eq.\ (\ref{Q}) will be termed the {\em quantum Chernoff overlap} of the states $\rho$ and $\sigma$. Notice that $Q(\rho,\sigma)\leq 1$ 
and its maximal value is reached when the states $\rho$ and $\sigma$ 
coincide.

Let us denote by $||A||_1:={\rm Tr}|A|$ the trace norm of a trace-class operator $A$. Originally, Hillery employed the trace metric 
$2\,T(\rho,\sigma):=||\rho-\sigma||_1$ to define the nonclassical 
distance \cite{Hi}. In the symmetric particular case $s=\frac{1}{2}$,
Holevo proved the following pair of inequalities \cite{Hol}:
\begin{equation} 
1-Q_{1/2}(\rho,\sigma)\leq T(\rho,\sigma) \leq 
\sqrt{1-\left[Q_{1/2}(\rho,\sigma)\right]^2}. \label{4}
\end{equation}
 Equation \ (\ref{4}) shows that the trace $Q_{1/2}(\rho,\sigma)$ 
is a measure of distinguishability as good as the trace metric  
$2\,T(\rho,\sigma)$. Having $Q(\rho,\sigma) \leq Q_{1/2}(\rho,\sigma),$ 
Holevo's second inequality in Eq.\ (\ref{4}) provides the upper bound  
of the Chernoff overlap $Q(\rho,\sigma)$ in terms of the trace distance, 
$Q^2\leq 1-T^2$. The lower bound $Q\geq 1-T$ 
was proven in Refs.\cite{ch2}, so that the inequalities \ (\ref{4}) 
still hold when replacing $Q_{1/2}(\rho,\sigma)$ by $Q(\rho,\sigma)$. 
We write them down below together with some other properties of the functions $Q_s$ and $Q$ proven in Refs.\cite{ch2}:

\begin{enumerate}
\item {\em Relations to the trace distance:}
$$ 1-Q(\rho,\sigma)\leq T(\rho,\sigma) \leq 
\sqrt{1-\left[Q(\rho,\sigma)\right]^2}.$$
\item {\em Convexity in $s$ of the trace $Q_s(\rho,\sigma)$}, 
Eq.\ (\ref{Renyi}). As a consequence, $Q(\rho,\sigma)$  is the unique
minimum of the R\'enyi overlaps $Q_s(\rho,\sigma)$.
\item {\em Joint concavity in $\{\rho,\sigma\}$ of the R\'enyi 
overlaps $Q_s$ and of their lower bound $Q$.} This means that 
the R\'enyi overlaps between a given state, say $\rho$, 
and an arbitrary set of states display a unique maximum.
\item {\em Multiplicativity of $Q_s$,}
$$ Q_s(\rho_1\otimes\rho_2,\sigma_1\otimes\sigma_2)
=Q_s (\rho_1,\sigma_1)Q_s (\rho_2,\sigma_2),$$
which is equivalent to the identity:
$$ {\rm Tr}\left [(\rho_1\otimes\rho_2)^s (\sigma_1\otimes\sigma_2)^{1-s}
\right]= {\rm Tr}(\rho_1^s\otimes\sigma_1^{1-s})\;{\rm Tr}(\rho_2^s
\otimes\sigma_2^{1-s}).$$
\item {\em Invariance of $Q(\rho,\sigma)$ under unitary transformations.}
\item {\em Monotonic increase of $Q(\rho,\sigma)$ 
under completely positive, trace-preserving maps.}
\end{enumerate}

In Refs.\cite{ch1,Kar} the relation between the Chernoff overlap
and the Uhlmann fidelity is largely discussed. Recall that Uhlmann  introduced the fidelity ${\cal F}(\rho, \sigma)$ of two mixed states, 
$\rho$ and $\sigma$, as the maximal quantum-mechanical transition probability between {\em all} purifications  of the given states 
\cite{Uhl}. Uhlmann wrote the distance $d_{B}$ between two states
discovered by Bures \cite{Bu} in terms of the transition probability 
${\cal F}$ as $\left[d_{B}(\rho,\sigma)\right]^2
=2[1-\sqrt{{\cal F}(\rho,\sigma)}],$ 
and succeeded in finding an explicit expression of the fidelity 
\cite{com}: 
\begin{eqnarray}
{\cal F}(\rho, \sigma)=\left[{\rm Tr}\left(\sqrt{\sqrt{\rho}\;\sigma
\sqrt{\rho}}\right)\right]^2. \label{ex} 
\end{eqnarray}
The above formula, ${\cal F}(\rho, \sigma)=(||\sqrt{\rho} 
\sqrt{\sigma}||_1)^2$, combined with a result of Fuchs and 
van de Graaf in Ref.\cite{FG}  provides two important bounds: 
\begin{equation}
{\cal F}(\rho, \sigma)\leq Q(\rho, \sigma)\leq 
\sqrt{{\cal F}(\rho, \sigma)}. \label{FG}
\end{equation}
When one of the states is pure, $Q(\rho, \sigma)$ equals the fidelity 
\cite{Kar} and has thus the significance of a transition probability: 
$Q(\rho, \sigma)= {\cal F}(\rho, \sigma)={\rm Tr} (\rho \sigma).$
$\sqrt{{\cal F}}$ shares with  $Q_s$ and $Q$ the properties 
1 and 3-6, which are precisely the demands for a genuine measure 
of nonclassicality, stated in Ref.\cite{PTH1} . In view of these 
properties, we introduce here an ideal Chernoff degree of
nonclassicality for an arbitrary state $\rho$:
\begin{eqnarray}
D^{(C)}(\rho):=\min_{{\rho}^{\prime} \in {\cal C}}[1-Q(\rho,{\rho}^{\prime})], \label{deg}
\end{eqnarray}
where ${\cal C}$ is the set of all classical states. According to
Eq.\ (\ref{Q}), the Chernoff degree of nonclassicality\ (\ref{deg}) 
vanishes when the given state $\rho$ is classical.  Definition 
\ (\ref{deg}) implies the maximization of the Chernoff overlap
$Q(\rho, {\rho}^{\prime})$ over the whole set of classical states 
${\rho}^{\prime}\in {\cal C}$. 

We will focus on the nonclassicality of single-mode Gaussian 
states of the radiation field, which are especially useful 
in experiments. Taking advantage of their simple parametrization, 
we apply the definition \ (\ref{deg}) to evaluate a Gaussian 
degree of nonclassicality, just as in Refs.\cite{PTH1,PTH2}. 
Such a Gaussian approach consists in replacing the reference set 
${\cal C}$ of all classical one-mode states by its subset 
${\cal C}_0$ consisting only of Gaussian ones. 

Recall that any one-mode Gaussian state $\rho_G$ can be parametrized 
as a displaced squeezed thermal state (DSTS) \cite{PT1}:
\begin{equation}
\rho_G=D(\alpha) S(r,\varphi) \rho_T S^{\dag}(r,\varphi) 
D^{\dag}(\alpha). \label{dst}
\end{equation}
Here $D(\alpha)=\exp{(\alpha a^{\dag}-\alpha^* a)}$ is a Weyl 
displacement operator with the coherent amplitude $\alpha$,
$S(r,\varphi)=\exp{\{\frac{1}{2}  r[{\rm e}^{i\varphi} (a^{\dag})^2
-{\rm e}^{-i\varphi} a^2]\}}$ 
is a Stoler squeeze operator with the squeeze factor $r$ 
and squeeze angle  $\varphi$, and
\begin{equation}
\rho_{T}=\frac{1}{\bar{n}+1}\sum_{n=0}^{\infty} 
\left(\frac{\bar{n}}{\bar{n}+1}\right)^n|n \rangle \langle n|
\label{to} 
\end{equation}
is a thermal state with the Bose-Einstein mean occupancy 
$\bar{n}=[\exp(\beta\hbar\omega)-1]^{-1}$. 

We consider a nonclassical single-mode Gaussian state $\rho_G$  whose parameters are: $\alpha,\; \varphi,\; \bar n$, 
and $r>r_c:=\frac{1}{2}\ln(2\bar{n}+1)$. Any classical one-mode state 
${\rho}_G^{\prime} \in {\cal C}_0$  is identified by its
parameters $\alpha^{\prime},\; \varphi^{\prime},\; \bar{n}^{\prime},$ 
and $r^{\prime}$ subjected to the classicality condition 
$ r^{\prime}\leq r_c^{\prime}:=\frac{1}{2}\ln(2\bar{n}^{\prime}+1)$. 
When employing Eq.\ (\ref{deg}), the  Gaussian degree 
of nonclassicality of the state $\rho_G$ reads:
\begin{eqnarray}
D^{(C)}_0(\rho_G):=1-\max_{{\rho}_G^{\prime} \in {\cal C}_0} 
Q(\rho_G,\;{\rho}_G^{\prime}). \label{deg0}
\end{eqnarray}
We evaluate the R\'enyi overlap\ (\ref{Renyi}) of the pair of states
$\{{\rho}_G,\;{\rho}_G^{\prime}\}$ as a Hilbert-Schmidt scalar product:
\begin{equation}
Q_s(\rho_G,\;{\rho}_G^{\prime})=\frac{1}{\pi}\int d^2\lambda 
\chi_G^*(s, \lambda)\chi_G^{\prime}(1-s, \lambda). \label{RHS}
\end{equation}
In Eq.\ (\ref{RHS}), $\chi_G(s, \lambda)$ and 
$\chi_G^{\prime}(1-s, \lambda)$ are the weight functions in the Weyl
expansions of the one-mode Gaussian operators $(\rho_G)^s$ and 
$(\rho_G^{\prime})^{1-s}$, respectively. We write down the explicit
expression
\begin{eqnarray}
\chi_G(s, \lambda)&=&\frac{1}{(\bar n+1)^s-{\bar n}^s}
\exp{\left\{-\left[f(s, \bar n)+\frac{1}{2}\right]|\tilde \lambda|^2\right\}}
\nonumber 
\\&&\times \exp{\left[\lambda {\alpha}^*-{\lambda}^*\alpha \right]}, \label{chi}
\end{eqnarray}
with the notations:
\begin{eqnarray*}f(s, \bar n):=\frac{{\bar n}^s}{(\bar n+1)^s-{\bar n}^s}, 
\;\;\; (0<s<1);\end{eqnarray*}
\begin{eqnarray*}\tilde \lambda:=\cosh(r)\lambda-{\rm e}^{i\varphi}\sinh(r){\lambda}^*.\end{eqnarray*}
As shown in Ref.\cite{PT1}, Appendix A, the Gaussian integral 
in the r. h. s. of Eq.\ (\ref{RHS}) can be readily performed to get
the formula
\begin{eqnarray}
Q_s(\rho_G,\; {\rho}_G^{\prime})&=&\frac{1}{{\bar n}^s
({\bar n}^{\prime})^{1-s}}f(s, \bar n)f(1-s, {\bar n}^{\prime})
\frac{1}{\sqrt{K^2-|L|^2}}\nonumber 
\\&& \times
\exp{\left[-\frac{K|M|^2+{\Re e}(L^*M^2)}{K^2-|L|^2}\right]},
\label{Q_s}
\end{eqnarray}
where we have denoted:
\begin{eqnarray*}K:=\left[f(s, \bar n)+\frac{1}{2}\right]\cosh(2r)
+\left[f(1-s, {\bar n}^{\prime})+\frac{1}{2}\right]\cosh(2r^{\prime}), 
\end{eqnarray*}
\begin{eqnarray*}&&L:=\left[f(s, \bar n)+\frac{1}{2}\right]{\rm e}^{i\varphi} \sinh(2r)\nonumber \\&&
+\left[f(1-s, {\bar n}^{\prime})+\frac{1}{2}\right]
{\rm e}^{i{\varphi}^{\prime}} \sinh(2r^{\prime}),\end{eqnarray*}
\begin{eqnarray}M:={\alpha}^{\prime}-\alpha.  \label{KLM}
\end{eqnarray}
We mention that an equivalent expression of the R\'enyi overlap 
$Q_s({\rho}_G, {\rho}_G^{\prime})$ has already been found 
in Ref.\cite{ch3} as Eq. (91) thereof.
Maximization of $Q_s({\rho}_G, {\rho}_G^{\prime})$, Eqs.\ (\ref{Q_s})  
and\ (\ref{KLM}), with respect to the displacement $\alpha^{\prime}$ 
and the squeeze angle $\varphi^{\prime}$ yields obvious  values of these parameters for the closest classical state $\tilde {\rho}_G^{\prime}$:  
$\tilde {\alpha}^{\prime}=\alpha$ and $\tilde{\varphi}^{\prime}=\varphi$.
It turns out that nonclassicality is invariant under classical operations such as translations and rotations. A natural assumption is that 
$\tilde {\rho}_G^{\prime}$ belongs to the boundary of the set 
${\cal C}_0$ of all classical one-mode Gaussian states:
$\tilde {\rho}_G^{\prime} \in \partial {\cal C}_0.$ This means that 
the closest classical state $\tilde {\rho}_G^{\prime}$ is 
at the {\em classicality threshold} specified by the condition 
$
r^{\prime}=r_c^{\prime}:=\frac{1}{2}\ln(2\bar{n}^{\prime}+1)
\Longleftrightarrow \bar{n}^{\prime}={\rm e}^{r^{\prime}}
\sinh(r^{\prime}). $
After introducing all these findings into Eqs.\ (\ref{Q_s}) 
and\ (\ref{KLM}), the R\'enyi overlap 
$Q_s({\rho}_G, {\rho}_G^{\prime})$ becomes a two-variable function:
\begin{eqnarray}Q_G(s,r^{\prime})&=&\left\{\left[(\bar n+1)^s\sinh^{1-s}(r^{\prime})-\bar{n}^s\cosh^{1-s}(r^{\prime})\right]^2\right.\nonumber \\&&
\left.+\left[\cosh^{2(1-s)}(r^{\prime})-\sinh^{2(1-s)}(r^{\prime})
\right]\right.\nonumber \\&&
\left.\times \left[(\bar n+1)^{2 s}-\bar{n}^{2 s}\right]\cosh^2 (r-r^{\prime})\right\}^{-1/2}\nonumber \\&& \times{\exp}{[-(1-s) r^{\prime}]}.
\label{qs}\end{eqnarray}
Let us denote by ${\tilde Q}_G$ the maximum of the Chernoff overlap
in the r. h. s. of Eq.\ (\ref{deg0}):
\begin{equation}
{\tilde Q}_G:=\max_{{\rho}_G^{\prime} \in {\cal C}_0}
\min_{0\leq s \leq 1}Q_s(\rho_G,{\rho}_G^{\prime})
=\min_{0\leq s \leq 1}\max_{{\rho}_G^{\prime} \in {\cal C}_0}
Q_s(\rho_G,{\rho}_G^{\prime}). \label{saddle}
\end{equation}
We aim to find the value ${\tilde Q}_G$ of the function\ (\ref{qs})
that is reached for a pair of optimal values of its variables, 
hereafter denoted by $\tilde{s}$ and $\tilde{r}^{\prime}:\; 
{\tilde Q}_G=Q_G(\tilde{s}, \tilde{r}^{\prime})$. An analytic solution
can be found only if $\rho_G$ is a pure state, namely, a displaced
squeezed vacuum state. Indeed, when setting $\bar n=0$ 
into Eq.\ (\ref{qs}), we readily get the corresponding solution:
$
\tilde{s}=0,\;\;\; \tilde{r}^{\prime}=0,\;\;\; {\tilde Q}_G={\rm sech}(r). $
For any mixed nonclassical state $\rho_G$, the optimal parameters 
$\tilde{s}$ and $\tilde {r}^{\prime}$ cannot be determined 
analytically. This situation is similar to that encountered 
when using the relative entropy as a measure of nonclassicality 
for single-mode Gaussian states in Ref.\cite{PTH2}. According to 
Eq.\ (\ref{saddle}), ${\tilde Q}_G=Q_G(\tilde{s}, \tilde{r}^{\prime})$ 
is a saddle point of the function\ (\ref{qs}).

\begin{figure}[ht]
\includegraphics[width=5.5cm,angle=0]{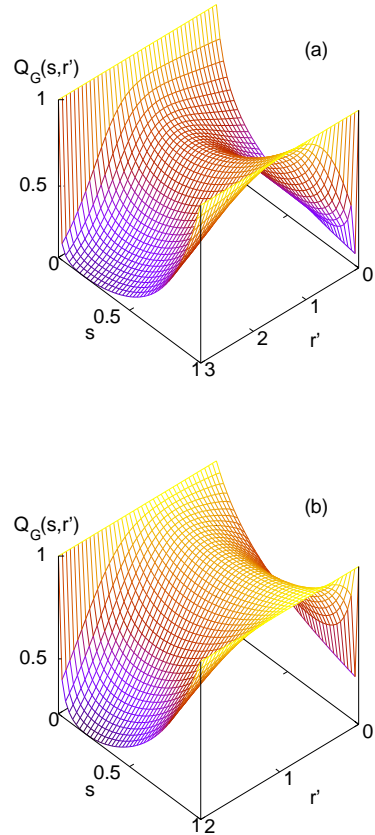}
\caption{(Color online) Displaying the saddle-point evaluations. 
The function $Q_G(s,r^{\prime})$, Eq.\ (\ref{qs}),
is plotted versus $r^{\prime}$ and $s$ 
for two nonclassical Gaussian states having the parameters: 
a) $r=2,\bar n=1,$ for which we get the saddle-point values: 
${\tilde Q}_G=0.617,\; \tilde{r}^{\prime}=1.265,\; \tilde s=0.283$; 
b) $r=1, \bar n=0.5,$ with the saddle-point results: 
${\tilde Q}_G=0.893,\; \tilde{r}^{\prime}=0.653,\; \tilde s=0.387$.}
\label{fig1}
\end{figure}

 The saddle-point numerical results can be seen in Fig.\ref{fig1}, where the  R\'enyi 
overlaps $Q_G(s, r^{\prime})$ are plotted versus the variables $s$ 
and $r^{\prime}$. 

It is now interesting to compare the present results with similar ones, found previously by using the Bures metric to quantify 
the nonclassicality of one-mode Gaussian states. In Ref.\cite{PTH1}, 
a Gaussian degree of nonclassicality has been defined as follows:
\begin{equation}
D^{(B)}_0(\rho_G):=1- \max_{{\rho}_G^{\prime} \in {\cal C}_0}
\sqrt{{\cal F}(\rho_G,{\rho}_G^{\prime})}. \label{Bures}
\end{equation}
Maximization of the fidelity could be performed analytically 
to give the simple result
\begin{equation}
\tilde{\cal F}:=\max_{{\rho}_G^{\prime} \in {\cal C}_0}
{\cal F}(\rho_G,{\rho}_G^{\prime})={\rm sech} (r-r_c). \label{fidmax}
\end{equation}

\begin{figure}[h]
\includegraphics[width=5.5cm,angle=0]{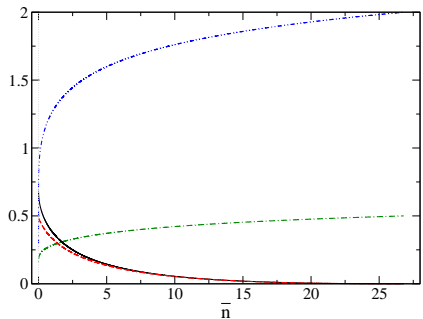}
\caption{(Color online) Chernoff degree of nonclassicality $D^{(C)}_0$ 
(full line), Eq.\ (\ref{deg0}), Bures degree $D^{(B)}_0$ (dashed line), 
Eq.\ (\ref{Bures}), versus the thermal mean occupancy $\bar n$ for nonclassical states with the squeeze factor $r=2.$ 
We have also plotted the corresponding saddle-point values  
$\tilde{r}^{\prime}$ (double-dot dashed line) 
and $\tilde{s}$ (dot-dashed line).} \label{fig2}
\end{figure}

\begin{figure}[h]
\includegraphics[width=5.5cm,angle=0]{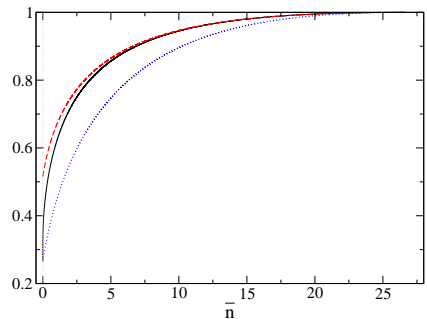}
\caption{(Color online) Displaying the inequalities \ (\ref{FG}) 
for the saddle-point value ${\tilde Q}_G$. Variation 
of the  optimal Chernoff value ${\tilde Q}_G$ (full line), 
the optimal fidelity $\tilde{\cal F}$ (dotted line) 
and $\sqrt{\tilde{\cal F}}$ (dashed line) with the mixedness 
of the nonclassical Gaussian state $\rho_G$. The squeeze parameter 
is $r=2$.} \label{fig3}
\end{figure}

We present in Fig. \ref{fig2} the degrees of nonclassicality 
\ (\ref{deg0}) and \ (\ref{Bures}) as  functions of the mixedness 
parameter $\bar n$  at a fixed squeeze factor $r>0$. They have 
close graphs over the whole nonclassicality domain of the squeezed 
thermal state $\rho_G$: $0\leq \bar n<{\bar n}_c:={\rm e}^r\sinh(r)$. 
Also plotted are the corresponding saddle-point parameters $\tilde{s}$ 
and $\tilde{r}^{\prime}$. Both of them are increasing functions of the
variable $\bar n$, starting from the pure-state values  $\tilde{s}=0$ and $\tilde{r}^{\prime}=0$,
and ending at the threshold values $\tilde{s}=\frac{1}{2}$ 
and $\tilde{r}^{\prime}=r$, respectively. Making use of Eq.\ (\ref{qs}),
we have proven that if $r=r_c^{\prime}$, then the optimal $\tilde{s}$ tends to $\frac{1}{2}$. The curves in Fig. \ref{fig2}  
are in fact calibration graphs for an easy reading 
of the nonclassicality properties of the squeezed mixed state $\rho_G$.
In addition, the bounds in Eq.\ (\ref{FG}) are illustrated 
in Fig. \ref{fig3} by plots of the optimal values ${\tilde Q}_G,\; 
\tilde{\cal F},$ and $\sqrt{\tilde{\cal F}}$ versus the thermal mean occupancy $\bar n$ of the nonclassical state $\rho_G$. 
Inequalities 
\ (\ref{FG}) are clearly displayed in this figure: ${\tilde Q}_G$ 
coincides with $\tilde{\cal F}$ for pure states ($\bar n=0$) 
and becomes rather close to $\sqrt{\tilde{\cal F}}$ as the degree 
of mixing increases. 

To conclude, in this work we have shown that the remarkable properties of quantum Chernoff bound can be used to discriminate 
between a (nonclassical) state and a set of (classical) states. 
 We have chosen the class   
of one-mode Gaussian states, for which an explicit expression 
of the R\'enyi overlap is recovered as Eqs.\ (\ref{Q_s}) 
and \ (\ref{KLM}). In general, the Chernoff overlap
$Q(\rho_G,\;{\rho}_G^{\prime})$ could be computed only numerically, 
while analytic expressions of the corresponding fidelity 
${\cal F}(\rho_G,\;{\rho}_G^{\prime})$ are at hand for a long time 
 \cite {fid1G}. 
However,the numerical calculation of the Chernoff degree of nonclassicality by saddle-point methods is straightforward 
and can be performed with great accuracy.
 Our present results 
are consistent with the analogous ones obtained previously by use 
of the Bures metric, in accordance with the general relations 
between the Chernoff overlap and the Uhlmann fidelity. 

This work was supported  by the Romanian 
Ministry of Education and Research through Grant No. IDEI-995/2007 for 
the University of Bucharest.


\begin{thebibliography}{99}
\bibitem{com} We have chosen to preserve the physical significance 
of the fidelity as a transition probability between optimal 
purifications. In Refs.\cite{ch1,Fuchs,Kar}, fidelity is defined 
as the square root of Uhlmann's fidelity,  Eq.\ (\ref{ex}).
\end{thebibliography}
\end{document}